\documentclass[traditabstract]{aa} 
\usepackage[varg]{txfonts}
\usepackage{time}
\usepackage{color}
\usepackage{natbib}
\usepackage[percent]{overpic}
\usepackage{array}
\newcolumntype{C}{>{$}c<{$}} 
\usepackage{amstext,amsmath,mathtools}
\usepackage{graphicx}
\usepackage{textcomp}
\usepackage{txfonts}
\usepackage{url}
\usepackage{booktabs} 
\begin{document}
\titlerunning{Improved high--order aberration compensation with PD and MFBD image restoration}
\title{Improved compensation for missing high-order aberrations with multi-frame blind deconvolution and phase diversity techniques}

\author{G.B. Scharmer\inst{1,2}
}

\institute{Institute for Solar Physics, Dept. of Astronomy, Stockholm University,
AlbaNova University Center, SE 106\,91 Stockholm, Sweden \and
EST Foundation, C/ Via Láctea s/n 38200 La Laguna, Spain 
}
\date{Draft: \now\ \today}
\frenchspacing

\abstract
{Image restoration with phase diversity (PD) and multi-frame blind deconvolution (MFBD) techniques lead to remarkable improvement in image quality, but is limited by the number of aberration parameters that can be determined. Compensation for the missing high-order modes is needed to further improve image quality, and to compensate for residual straylight.}
{We aim to improve on methods for compensating missing high-order aberrations previously proposed by Scharmer et al. (2010) and L\"ofdahl \& Hillberg (2022).} 
{The wavefront is divided into a set of low-order aberrations, which are treated individually, and high-order aberrations that are treated collectively. The high-order part of the transfer function is averaged, assuming Gaussian seeing statistics with Kolmogorov turbulence for the individual aberrations. Assuming that their accumulated effect is roughly uniform over the pupil, we obtain a high-order modulation transfer function that is represented with a single free parameter.}    
{The high quality of the of the proposed high-order modulation transfer function is verified by calculating the Strehl from random samples of point-spread functions for $D/r_0$ ranging from 4 to 20, where $D$ is the telescope diameter and $r_0$ Fried's seeing parameter. The proposed approximation offers significant advantages over the earlier proposals.}  
{The proposed method of compensating for missing high-order aberrations in PD or MFBD image reconstruction can dramatically improve the fidelity of solar images. Variations of the seeing ($r_0$) within a data-set can be compensated for self consistently during PD or MFBD processing, but the average value of $r_0$ must be provided from an external source. The computational cost of implementing this method is marginal.}

\keywords{ Methods: observational -- Techniques: image processing -- Techniques: high angular resolution }

\maketitle

\section{Introduction}\label{Introduction}
Present-day meter-class and larger solar telescopes cannot provide data close to their diffraction limits without advanced single-conjugate adaptive optics \citep[SCAO; e.g.,][]{1998SPIE.3353...72R, 2000SPIE.4007..239S, 2020SPIE11448E..0TJ, 2024A&A...685A..32S} or (for the largest telescopes) multi-conjugate adaptive optics (MCAO) systems \citep{2026SPIE14150E..0JJ} that in real-time compensate for the small-scale phase errors caused by turbulent seeing. The particular challenges of  such AO\footnote {Here and in the following, we refer to SCAO and MCAO systems collectively as AO systems.} systems on solar telescopes are the generally poor daytime seeing, the need to operate at visible wavelengths and to use low-contrast solar granulation, which is poorly resolved with sub-apertures of less than about 8--10~cm diameter, as target of the wavefront sensor (WFS). This sets a lower limit to the scale at which such wavefront errors can be compensated for, unavoidably leaving significant residual phase errors at smaller scales. Even at the best sites, this makes solar AO performance marginal under prevailing seeing conditions. In addition, there remain residual phase errors also at larger scales, which are partly from the limited efficiency of the AO system used, and partly from the inadequate compensation, with or without MCAO, for the variation of such phase errors over the observed field-of-view (FOV).

To compensate for the residual aberrations left in images obtained with solar telescopes, short exposure times in combination with image restoration techniques are in use on a regular basis. These are of two types: speckle and related techniques that rely on a statistical description of turbulent seeing \citep{1970A&A.....6...85L, 1993A&A...268..374V, 2007ApOpt..46.8015W, 2008A&A...488..375W}, and techniques that have their origin in the phase diversity (PD) and the multi-frame blind deconvolution (MFBD) techniques \citep{1979SPIE..207...32G, 1992JOSAA...9.1072P, 1994A&AS..107..243L, 2002SPIE.4792..146L, 2005SoPh..228..191V}. The advantage of the PD and MFBD techniques is that they allow for arbitrary wavefront perturbations, and thus more easily can handle residual phase errors from imperfect AO compensation. However, this is also a weakness, because existing PD and MFBD techniques can only compensate for a limited number of aberrations, necessarily leading to residual phase errors at small scales. For the techniques implemented by collaborators of the present author \citep{2002SPIE.4792..146L, 2005SoPh..228..191V}, that limit presently seems to be on the order of 80-100 aberrations. Whereas this is an acceptable limitation for a telescope such as the Swedish 1-m Solar Telescope \citep{2003SPIE.4853..341S} in good seeing conditions, it is more of a problem in less favourable conditions, and definitely a major problem for the largest 4-m class solar telescopes that are already in operation -- the Daniel K. Inouye Solar Telescope \citep[DKIST;][]{2018SPIE10700E..0VW,2020SoPh..295..172R}, or in the planning -- the European Solar Telescope \citep[EST;][]{2022A&A...666A..21Q}. 

The challenge of operating AO on increasingly larger telescopes can be quantified by estimating the variance of the residual wavefront errors that remain after perfect compensation of $M$ independent aberrations with a telescope of diameter $D$ in seeing conditions characterised by the Fried parameter $r_0$ \citep{1966JOSA...56.1372F}:
\begin{equation}
  \sigma^2 = 0.34\, (D/r_0)^{5/3} M^{-5/6},
  \label{eq:1}
\end{equation}
valid for zonal correction \citep[for a discussion and overview, see][chapter 3]{2004aoa..book.....R}. For the variance of the residual wavefront errors to remain independent of the telescope diameter, therefore $M$ needs to scale as $D^2$. A somewhat more favourable scaling is obtained for compensation with Karhunen-Loève (KL) modes, $M \propto D^{1.81}$. Nevertheless, to compensate for the degrading effects of seeing, it is clear that a 4-m class telescope needs an AO system with over one order of magnitude larger number of actuators than for a 1-m telescope. In addition to that challenge, MFBD and PD techniques used to process such data also need to handle the effects of a similar number of aberrations, to fully mitigate the effects of residual wavefront errors that are left after closed-loop compensation with the AO system. 

Post-processing of images from 4-m class solar telescopes thus faces two challenges: on the one hand, there is a need to compensate for $\sim$1000 independent residual aberrations from the action of the AO system. On the other hand there is also a need to compensate for residual aberrations at even smaller scales. It is well-known from the operation of night-time AO systems  that compensating for a limited number of low-order modes leaves a point spread function (PSF) that for long exposures has a diffraction limited core with low peak transmission (low Strehl, $R$) plus a ``halo" consisting of strongly enhanced outer diffraction rings with an integrated transmission of $1-R$ \citep[e.g., ][]{1992ESOC...42..475C}. In earlier work \citep{2010A&A...521A..68S}, we have demonstrated that the above type of PSF delivers images of solar granulation that reproduces fine structure well but with enhanced straylight that lowers the overall contrast. To optimise the fidelity of solar science data, this straylight compensation therefore is of significant importance. 

In this paper, we adress the problem of compensating for missing high-order aberrations, which is of high relevance to the quality of data from both meter-class and 4-m class solar telescopes. The paper is organised as follows: In Sect. \ref{sec:prev_work} we define the problem and describe previous efforts to address this problem. We describe the new approach in Sect. \ref{sec:prop_method}, along with simulations that compare Strehl values from high-order compensated PSF's with the previous and present methods. In Sect. \ref{sec:comp_impl}, we comment on the practical implementation of the method, and in Sect. \ref{sec:discussion} we summarize the discussion. 

\section{Previous work}
\label{sec:prev_work}

Processing of images with PD and MFBD techniques most often relies on the additive white Gaussian noise model \citep{1979SPIE..207...32G, 1992JOSAA...9.1072P, 1994A&AS..107..243L, 2002SPIE.4792..146L, 2005SoPh..228..191V}, 
leading to the following estimate of the restored object $\hat F$ from the observed images $D_k$ and their associated estimated transfer functions $\hat T_k$ 
\begin{equation}
  {\hat F} = \sum\limits_{k=1}^{N} D_k {\hat T}_k^*\biggm/\sum\limits_{k=1}^{N}{|{\hat T}_k|^2},
  \label{eq:2}
\end{equation}
\citep{1992JOSAA...9.1072P}, where $N$ is the number of images in a dataset, $|{\hat T}_k|^2 = {\hat T}_k {\hat T}_k^*$
and ${\hat T}_k^*$ is the complex conjugate of ${\hat T}_k$. Ignoring noise, we note that the observed images are related to true object $F$ and the observed images as $D_k = F~T_k$, where $T_k$ are the true transfer functions, including the missing high-order terms in addition to the low-order terms contained in ${\hat T}_k$. Inserting this in Eq. \ref{eq:2}, we obtain
\begin{equation}
  {\hat F} = F \sum\limits_{k=1}^{N} T_k {\hat T}_k^* \biggm/
  \sum\limits_{k=1}^{N}{|{\hat T}_k|^2} 
  \label{eq:3}
\end{equation}
We note that this is a relation of the form
\begin{equation}
  {\hat F} = S F,
  \label{eq:4}
\end{equation}
i.e., in the form of a multiplication of the true object with a
transfer function,
\begin{equation}
  S = \sum\limits_{k=1}^{N} T_k {\hat T}_k^* \biggm/ \sum\limits_{k=1}^{N}{|{\hat T}_k|^2} .
  \label{eq:5}
\end{equation}
Since MFBD and PD techniques applied to solar data always involve 10 or more raw images for each processed frame, we proposed \citep{2010A&A...521A..68S} to compensate for the unknown high-order aberrations contained in $T_k$ by replacing this with a statistical average, based on the Kolmogoroff turbulence model, assuming that $r_0$ at the time of recording the data is known.  Thus we estimate $S$ as
\begin{equation}
  \hat S  = \sum\limits_{k=1}^{N} \langle T_k \rangle {\hat T}_k^*  
  \biggm/ \sum\limits_{k=1}^{N}{|{\hat T}_k|^2} ~~,
  \label{eq:6}
\end{equation}
where angular brackets, $\langle\ldots\rangle$, indicate an ensemble average over many independent realisations of high-order aberrations, taken separately for each of the $N$ transfer functions with their individual sets of low-order aberrations. This approach is in particular appropriate when used with aberrations modeled with KL functions, which have statistically independent amplitudes for seeing obeying Kolmogoroff statistics.

To estimate the effect of high-order aberrations in $T_k$, we thus add many random higher-order KL modes, with amplitudes given by Kolmogorov statistics, to the low-order modes contained in each of the ${\hat T}_k$, and average the transfer equation. The required measurements of $r_0$ at the time of recording the data are obtained from the SST AO system, which automatically logs $r_0$ measurements every second when the telescope is open \citep{2019A&A...626A..55S, 2024A&A...685A..32S}. 

Simulations demonstrate that the above approximation gives a very good compensation for the missing high-order terms, even with a small number of recorded images. In this way, the proposed approach can be said to optimally combine PD and MFBD processing of arbitrary combinations of low-order aberrations with a speckle-like approach to including high-order modes originating from seeing. In addition, $\hat S$ is simple to calculate.

However, the implementation of the above with PD and MFBD processing is computationally expensive. This is because each iteration of such processing not only requires the calculation of  $\langle T_k \rangle $, but also the derivatives of $\langle T_k \rangle $ with respect to all low-order aberrations contained in ${\hat T}_k$. Each of these derivatives should be calculated for several (many) random samples of high-order modes.

The alternative is to instead relegate the calculation of $\hat S$ to that of an extra post-processing step, once the PD or MFBD processing is complete. With this approach, compensation of missing high-order terms cannot impact on the actual image restoration process, which probably would have a positive effect by providing better overall fits to the observed data.

\subsection{The statistical diversity method} 
To eliminate the above down-side of the proposed method, \citet{2022A&A...668A.129L} realised that even a single random sample of high-order aberrations suffices to provide a meaningful estimate of $ \langle T_k \rangle$. This is, because many high-order KL aberrations actually contribute to a single statistical sample contained in $\hat S$, such that even a single such sample contain the essential features of $\hat S$. \citet{2022A&A...668A.129L} therefore implemented, what they refer to as statistical diversity (SD), a simple modification of existing MFBD and PD processing. This consists in adding a single tail of high-order aberrations, based on the assumed value of $r_0$, to each estimated set of low-order aberrations. These same tails are used throughout the entire processing of a particular PD or MFBD dataset, such that the accumulated wavefront resulting from all the high-order terms can be pre-calculated once and for all. With this approach, the extra effort of processing is negligible, whereas simulations suggest that the method should work very well. The method has not yet been fully tested on actual SST data, and is therefore not yet implemented as part of the standard image reconstruction pipeline for SST data.

\section{The proposed method}
\label{sec:prop_method}
The method proposed is entirely consistent with the form of compensation for missing high-order aberrations given in Eq. \ref{eq:6} \citep{2010A&A...521A..68S}, but goes much deeper by developing the mathematics needed to actually carry out the averaging of the high-order part of the transfer function indicated by $\langle ... \rangle$ in that equation. To do this, we take advantage of well-established properties of atmospheric seeing characterised with KL aberrations. These are 1) that the KL coefficients corresponding to seeing are statistically independent; 2) that the KL coefficients have gaussian distributions with zero mean, and 3) that the variance of the KL coefficients are those given by \citet{1976JOSA...66..207N}.

The optical transfer function $T_k$ of an imaging system is described by the usual autocorrelation of the complex pupil function, which we write as 
\begin{equation}
T_k(\lambda \nu) = T_k(\delta x)=\int_{-\infty}^{\infty} A(x) \, A(x-\delta x) \, P_k(x) \, P_k^*(x-\delta x) \,dx,
\label{eq:7}
\end{equation}
where $A(x)$ represents the circular pupil and $x$ the two spatial coordinates in the pupil plane. Note that the separation (vector)  $\delta x$ maps directly onto the corresponding 2D spatial frequency $\nu$ in the Fourier plane via $\delta x = \lambda \nu$, where $\lambda$ is the wavelength. $P_k$ and its complex conjugate $P_k^*$ contain the phase errors at the pupil and is written
\begin{equation}
P_k(x)=\exp [i\, \phi_k(x)]
\label{eq:8}
\end{equation}
We expand the phase in two sets of  KL polynomials,
\begin{equation}
\phi_k(x)=\sum_2^{n_1} c_{kn} \psi_n(x) + \sum_{n_1+1}^\infty c_{kn} \psi_n(x) \equiv \phi_{k1}+\phi_{k2}
  \label{eq:9}
\end{equation}
where $\psi_n$ are the KL polynomials and $c_{kn}$ the corresponding KL coefficients. This splits the wavefront into two parts, where $\phi_{k1}$ represents the low-order part of the wavefront that is compensated for in the usual way with PD or MFBD techniques. $\phi_{k2}$ is the high-order part of the wavefront that we wish to provide a statistical average of. We do this by taking into account the assumed Gaussian distributions of the KL coefficients. We also note that any given KL coefficient occurs both in $P_k(x)$ and $P_k^*(x-\delta x)$ - thus these two quantities must be simultaneously included in the calculations when taking statistical averages. We thus make the approximation that the impact of the high-order tail of thousands of aberrations can be approximated with its expected statistical average:
\begin{equation}
\begin{aligned}
S_k(x,\delta x) = \exp [i (\phi_{k2}(x)-\phi_{k2}(x-\delta x)] \\ \approx \bigg\langle \exp \{ \, i \sum_{n_1+1}^\infty c_{kn} [\psi_n(x)
- \psi_n(x - \delta x) ] \} \bigg \rangle
\end{aligned}
  \label{eq:10}
\end{equation}
where the average ($\langle ....\rangle$) is taken over the ensemble of random $c_{kn}$ coefficients with variances $(D/r_{0k})^{5/3} \sigma_n^2$, allowing $r_0$ to vary within a burst of images. Since the KL coefficients are statistically independent, this average can be calculated independently for the quantities involving each KL coefficient, such that the above expression simplifies to 
\begin{equation}
\hat S_k(x,\delta x) = \prod_{n_1+1}^\infty \big\langle \exp\{i\, [ c_{kn} (\psi_n(x) - \psi_n(x - \delta x) ]\} \big \rangle \equiv \prod_{n_1+1}^\infty S_{kn}(x,\delta x)
  \label{eq:11}
\end{equation}
Taking the average over a Gaussian distribution of $c_{kn}$ values with variances $\sigma_n$, 
$S_{kn}$ is given by
\begin{equation}
S_{kn}(x,\delta x) = \exp \{- (D/r_{0k})^{5/3} \sigma_n^2 [\psi_n(x)-\psi_n(x - \delta x)]^2/2.\}
  \label{eq:12}
\end{equation}
\citep{1978JOSA...68...78W} and \citep[][Sects. 2.3 and 3.3]{2004aoa..book.....R}, where $\sigma_n^2$ are the variances given by \citet{1976JOSA...66..207N}, valid for $D/r_0$=1 . We can now calculate $\hat S_k(x,\delta x)$ as
\begin{equation}
\hat S_k(x,\delta x) = \exp\{- (D/r_{0k})^{5/3} \sum _{n_1+1}^{\infty} \sigma_n^2 [\psi_n(x)-\psi_n(x - \delta x)]^2/2\}. 
\end{equation}
  \label{eq:13}
We can write 
\begin{equation}
\hat S_k(x,\delta x) = \exp[-\alpha_k K(x,\delta x)],  
  \label{eq:14}
\end{equation}
where
\begin{equation}
K(x,\delta x) = \sum _{n_1+1}^{\infty} \sigma_n^2 [\psi_n(x)-\psi_n(x - \delta x]^2/2.
\label{eq:15} 
\end{equation}
and
\begin{equation}
\alpha_k=(D/r_{0k})^{5/3}
  \label{eq:16}
\end{equation}

\subsection{Computation of $\hat S$}

We note that $K(x,\delta x)$ in Eq. \ref{eq:15} is the result of the summation of thousands of KL contributions, each sampling the pupil with peaks and valleys at various quasi-random locations. With reasonable choices of $n_1$, this sampling of the pupil will on the average be relatively uniform, such that $K(x,\delta x)$ will show only small and smooth variations over the pupil coordinate x. A particular case is when $n_1$=1, corresponding to statistics of the atmosphere itself. In this case, $K(x,\delta x)$ is completely independent of $x$, and approaches $\exp[-3.44(\lambda \,\, \delta x/r_0)^{5/3}]$, given already by \cite{1961wptm.book.....T}. Whereas the same independence of $K$ on $x$ cannot be assumed for our calculations with $n_1 >> 2$, it should nevertheless serve as a reasonable approximation. The important consequence of this is that $\hat S_k$ can be moved outside the integral appearing in Eq. \ref{eq:7}, such that $\hat S_k$, if appropriately averaged (see below), directly can be treated as an atmospheric modulation transfer function (MTF) that in an approximate way compensates for the high-order ($n > n_1$) seeing contributions, for which only their variances can be assumed to be known. This implies that the effect of the high-order KL aberrations is approximated with a convolution of the PSF representing the low-order KL aberrations with a "straylight PSF" obtained from a statistical average of the optical transfer function of the high-order terms. Below, we demonstrate with numerical calculations that this is actually a very good approximation. 

To calculate an appropriately averaged $\hat S_k$, we average $K(x,\delta x)$ over $x$ to take into account the (assumed small) variations over the pupil. This average of $K(x,\delta x)$ is calculated within the overlapping pupil area, as in Eq. \ref{eq:7}, but normalising to the overlapping area of the two pupils. In this way, we measure the average atmospheric response within the spatial frequencies transmitted by the pupil and from the same parts of the pupil (given by the overlapping areas of $A(x)$ and $A(x-\delta x) $) as used in MFBD processing:
 
 \begin{equation}
 \hat{K} (\lambda \, \nu) = \hat{K}(\delta x)=\int_{-\infty}^{\infty} A(x) \, A(x-\delta x) \, K(x,\delta x) \,dx \, /C(\delta x) , 
\label{eq:17}
 \end{equation}
 where
 \begin{equation}
 C(\delta x) = \int_{-\infty}^{\infty} A(x) \, A(x-\delta x) \,dx
 \label{eq:18}
 \end{equation}
is the overlapping area for each $\delta x$. We then calculate $\hat S_k$ as in Eq. \ref{eq:14}
\begin{equation}
\hat S_k(\nu) = \exp[-\alpha_k  \, \hat{K}(\nu) ] ,
\label{eq:19}
\end{equation}
where $\hat S$ is equivalent to that given in Eq. \ref{eq:6}. This expression, obtained  by ignoring the (small) variations of $K(x,\delta x)$ over $x$, leads to the following approximation for the exact transfer function, which includes both the low-order aberrations contained in $\hat{T_k}$ and the compensation for missing high-order aberrations that contribute mostly straylight and are contained in $\hat S_k$: 
\begin{equation}
T_k \approx \langle T_k \rangle = \hat{S_k} \, \hat{T_k}.
\label{eq:20}
\end{equation}
Equations \ref{eq:19} and \ref{eq:20}, finally, allow us to calculate $\langle T_k \rangle$. This is the quantity in Eq. \ref{eq:6} that \citet{2010A&A...521A..68S} proposed to calculate by randomly sampling many high-order KL aberrations, and averaging the corresponding transfer functions. By pre-calculating the average of $K$ as described above, once and for all, we can obtain $\hat S_k$ for any $r_0$ at very small computational cost, by modifying the single constant $\alpha_k$. This flexibility is of importance, since $r_0$ normally varies while recording complex datasets that may include scans through a spectral line with polarimetry. Another advantage of the above approximation is that the derivative of $\hat S_k$ with respect to $\alpha_k$ (needed for iterative improvements) can be trivially calculated. Including this form of  straylight computation with MFBD processing, therefore, only requires a small computational effort that is comparable to increasing the number of KL coefficients by one.

We note in passing, that if $r_0$ is the same for all images in a burst, then $\langle T_k \rangle = \hat{S} \, \hat{T_k}$, which when inserted in Eq. \ref{eq:6} results in $\hat S = \hat S$, which confirms that the present approach and that of  \citet{2010A&A...521A..68S} indeed are consistent. However, the present approach goes further by solving the problem of how to actually calculate $\hat S$ efficiently without drawing many random samples of KL coefficients, and to model its variation with $r_0$.

\begin{figure}[!t]
\centering
\includegraphics[angle=90, width=0.99\linewidth,clip] {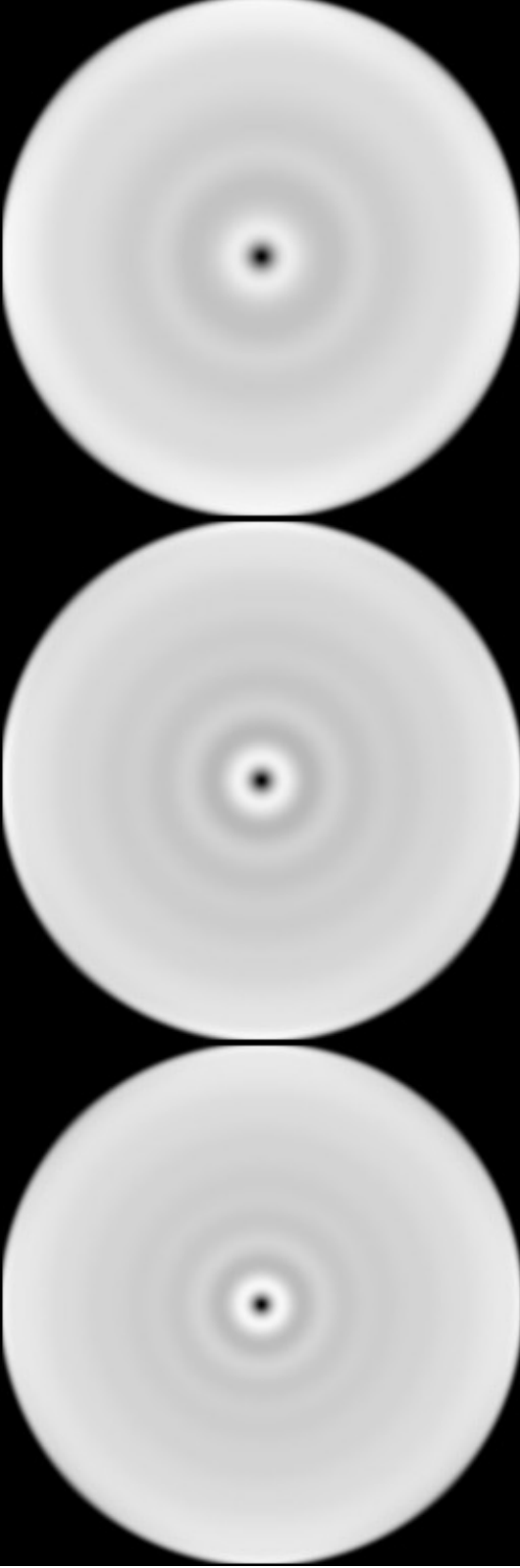}
\caption{The structure of $\hat {K}(\nu)$ as defined by Eq. \ref{eq:17}, for $n_1$=36 (left), 66 (middle), and 91 (right). The images are scaled individually. Note the virtual absence of azimuthal variations.}
 \label{fig:S-hat-comparison}
 \end{figure}

\begin{figure}[!t]
\centering
\includegraphics[angle=0, width=0.64\linewidth,clip] {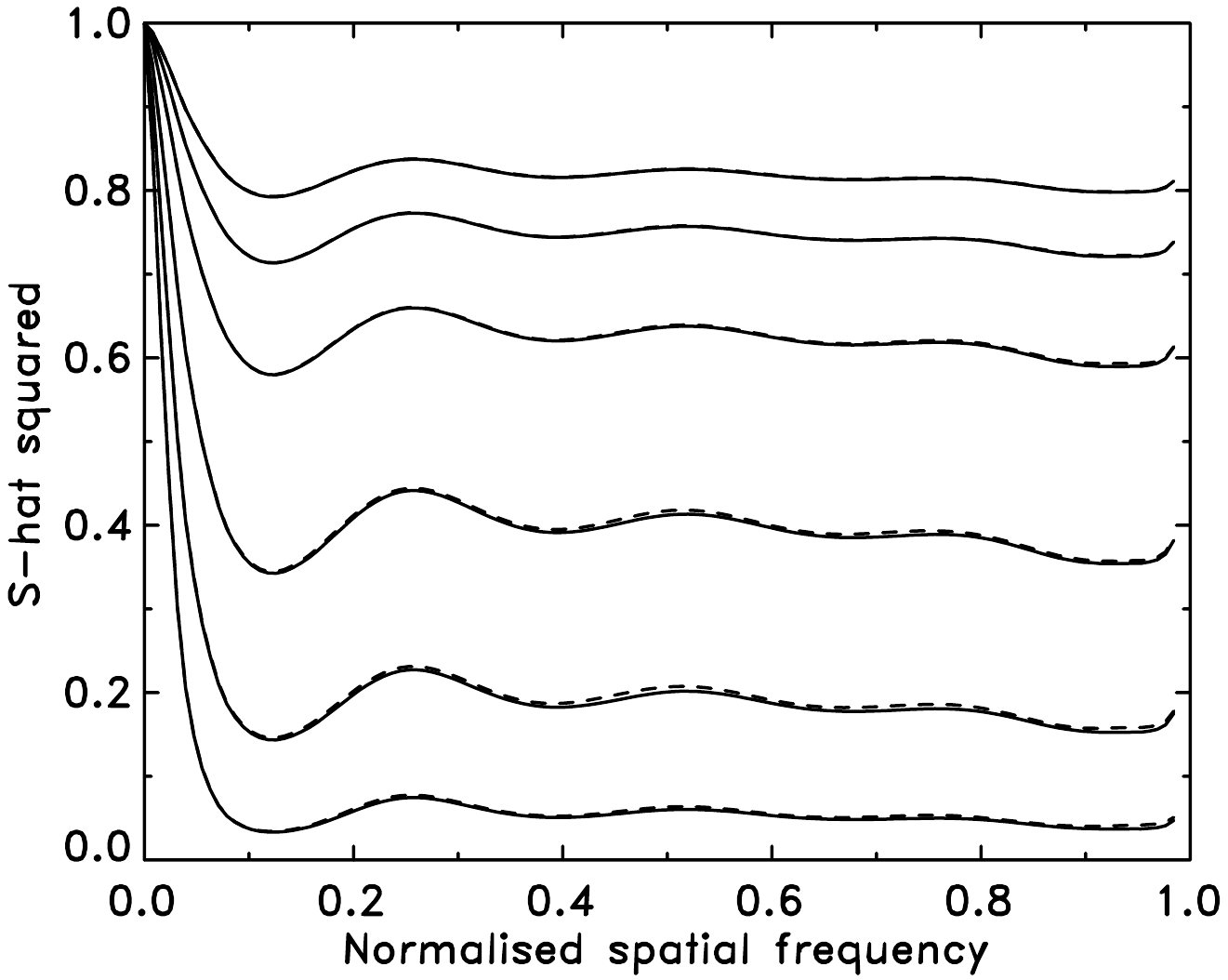}
\includegraphics[angle=0, width=0.64\linewidth,clip] {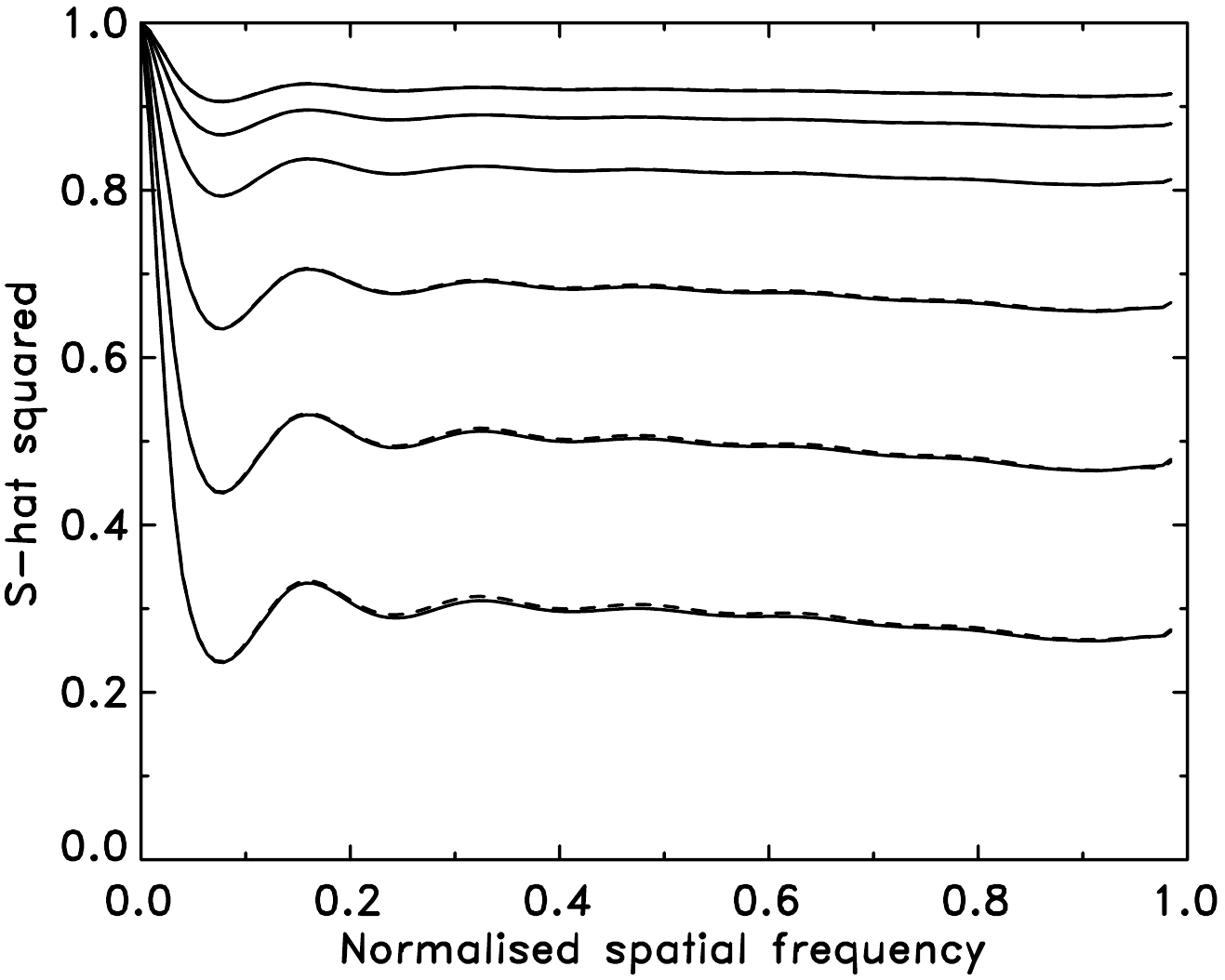}
 \caption{The azimuthal average of the square of $\hat S^2$, as calculated with the method proposed by \citet{2010A&A...521A..68S} (dashed, see Eq. \ref{eq:6}), and the method proposed here (Eq. \ref{eq:19}). We used $n_1$=36 (top) and $n_1$=91 (bottom), and included all KL aberrations up to $n$=2016. With the 2010 method, we used 4000 random sets of KL aberrations to obtain an accurate average. The curves are for a 1-m telescope with $r_0$=25, 20, 15, 10, 7, 5 cm (top to bottom). } 
 \label{fig:S-hat-comparison}
 \end{figure}

Figure \ref{fig:S-hat-comparison} shows a comparison between the methods for calculation of $\hat S$ of  \citet{2010A&A...521A..68S} and the method proposed here. Calculations were made for $D/r_0$ ranging from 4 to 20. It is evident that the two methods of calculation are equivalent.

The importance of the proposed method for calculating $\hat S$ is that its calculation only involves statistical averages of  high-order aberrations that are not compensated for with PD or MFBD image reconstruction. Therefore, compensation for the straylight caused by high-order aberrations can be done simultaneously with MFBD processing of the low-order modes without complex computational cross-talk between the two types of processing. 

\begin{figure*}[!t]
\centering
\includegraphics[angle=0, width=0.3\linewidth,clip] {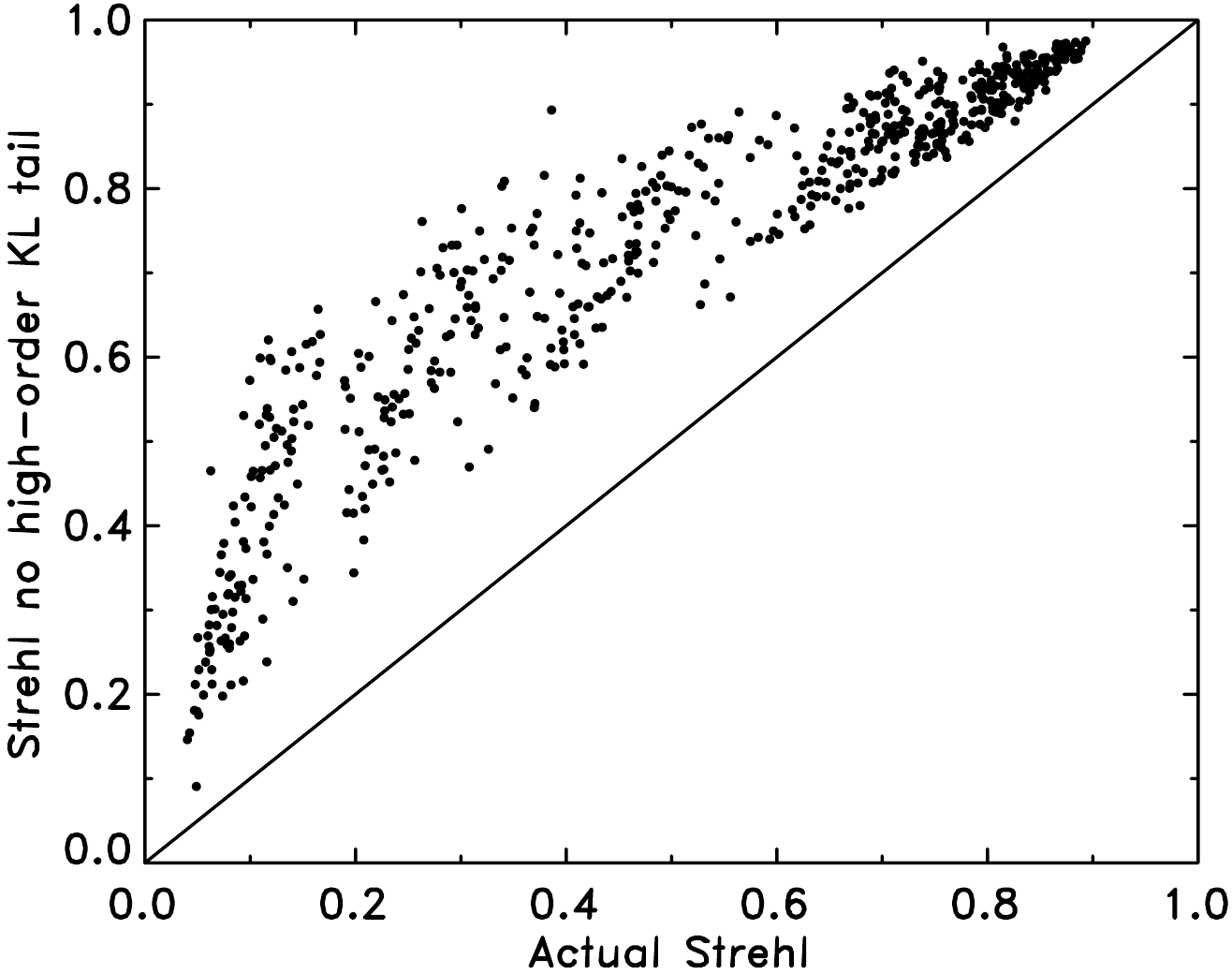}
\includegraphics[angle=0, width=0.3\linewidth,clip] {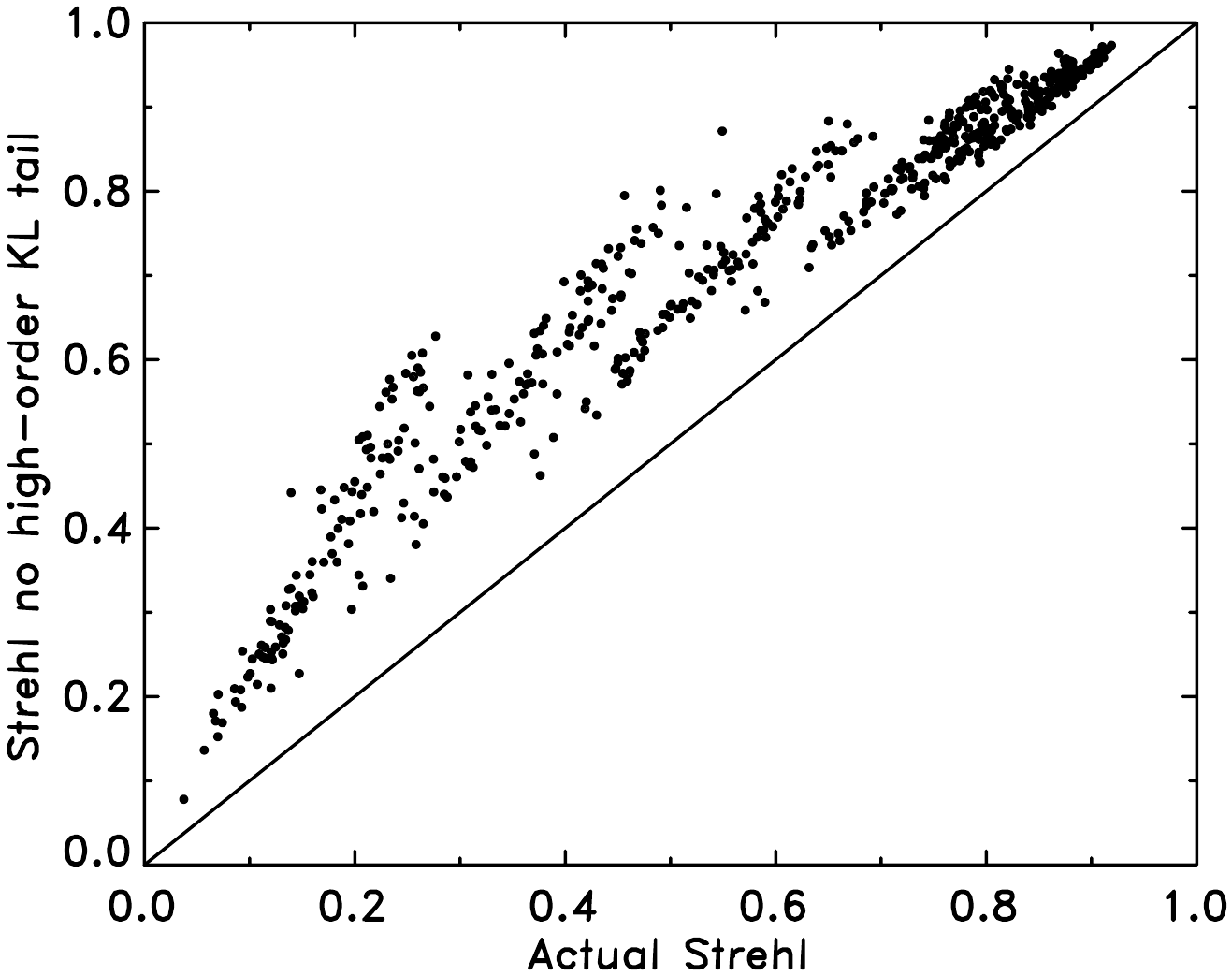}
\includegraphics[angle=0, width=0.3\linewidth,clip] {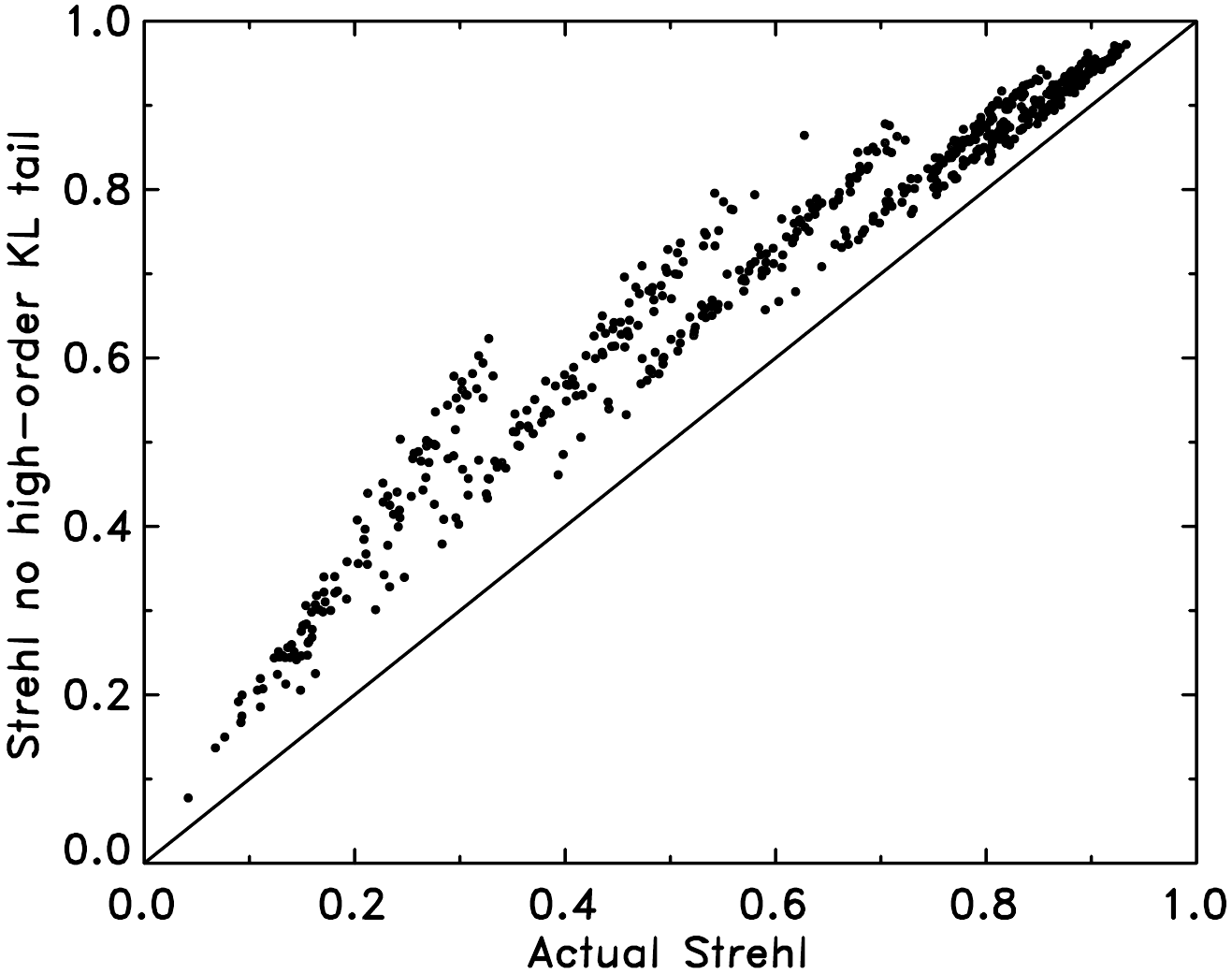}
\includegraphics[angle=0, width=0.3\linewidth,clip] {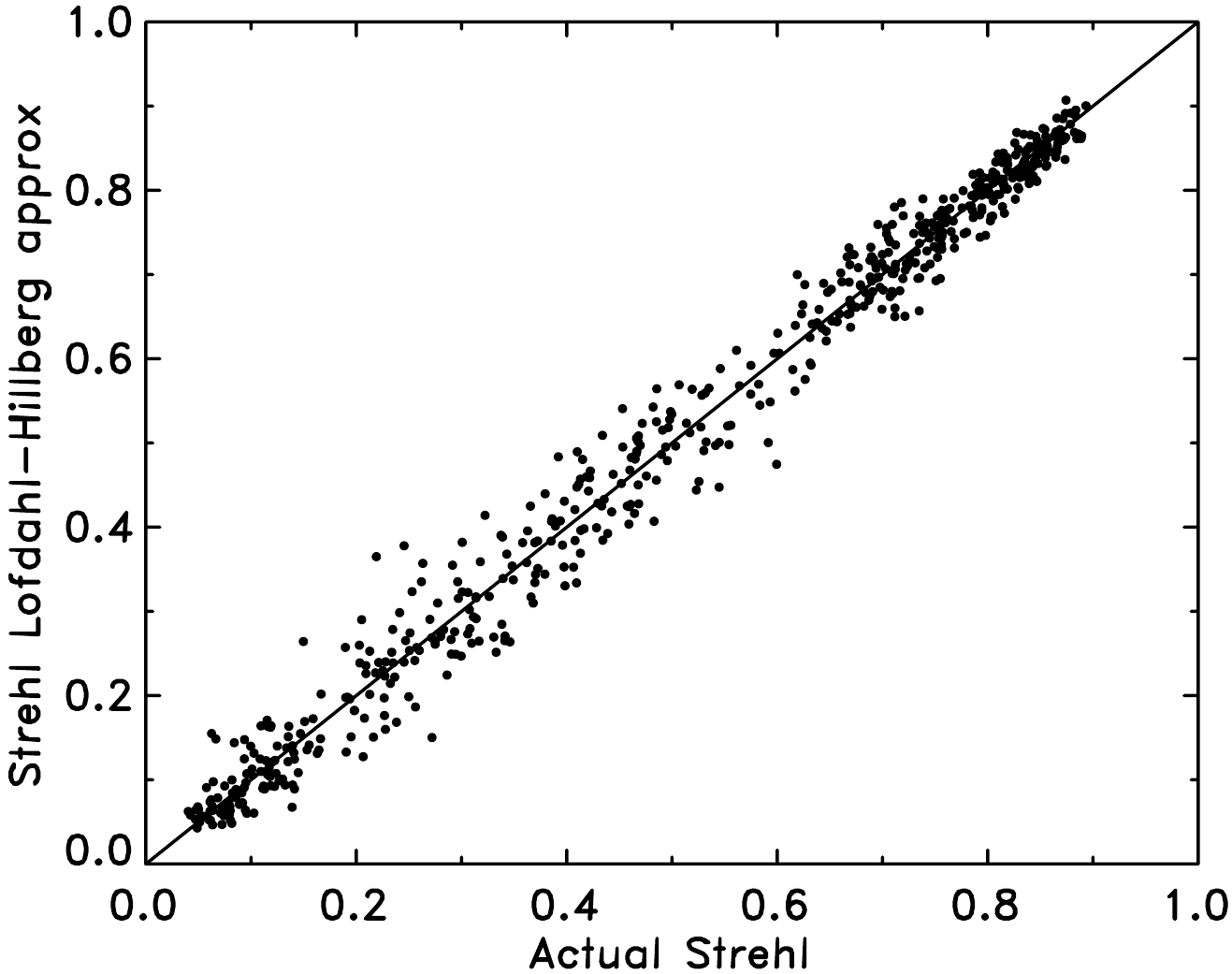 }
\includegraphics[angle=0, width=0.3\linewidth,clip] {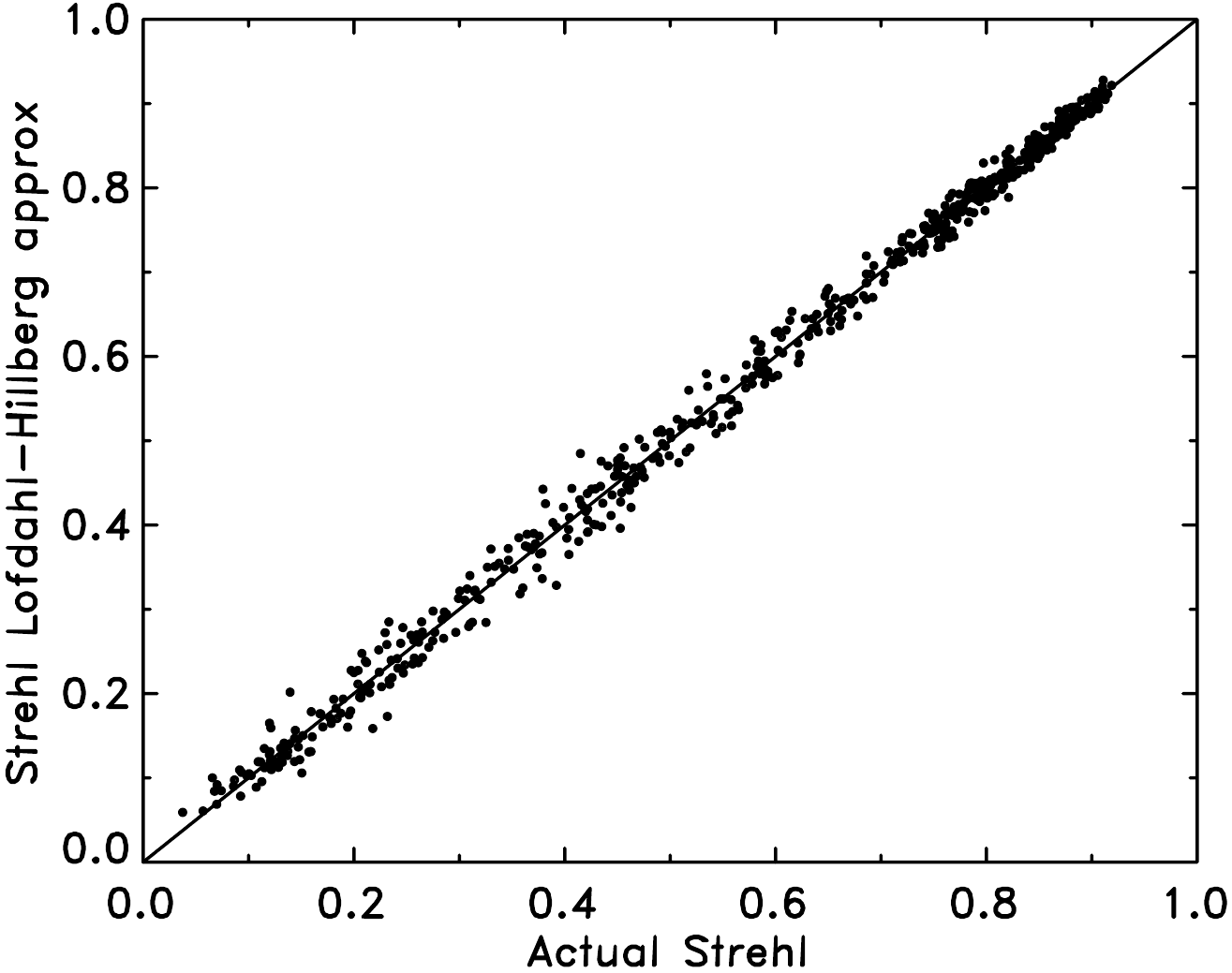 }
\includegraphics[angle=0, width=0.3\linewidth,clip] {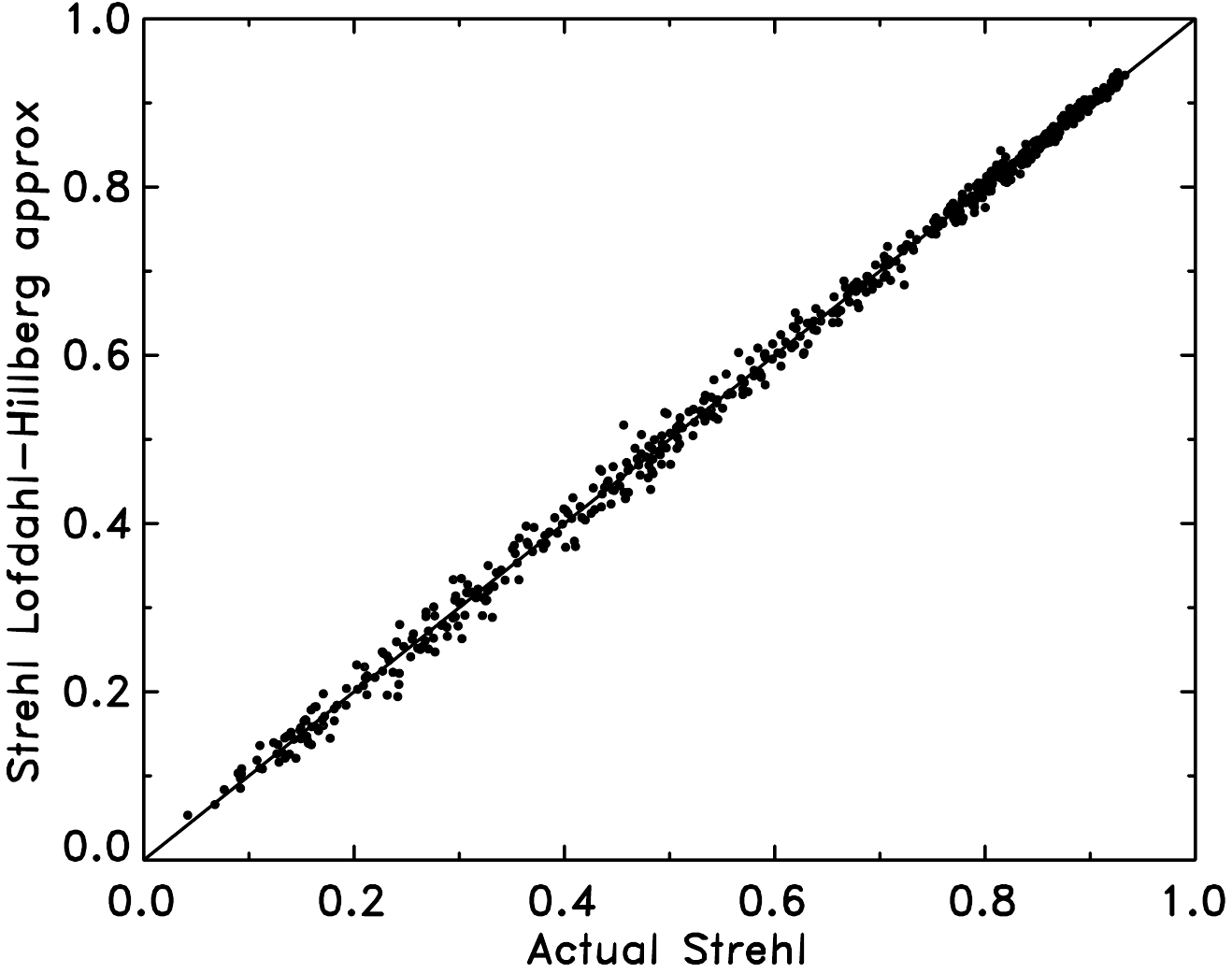 }
\includegraphics[angle=0, width=0.3\linewidth,clip] {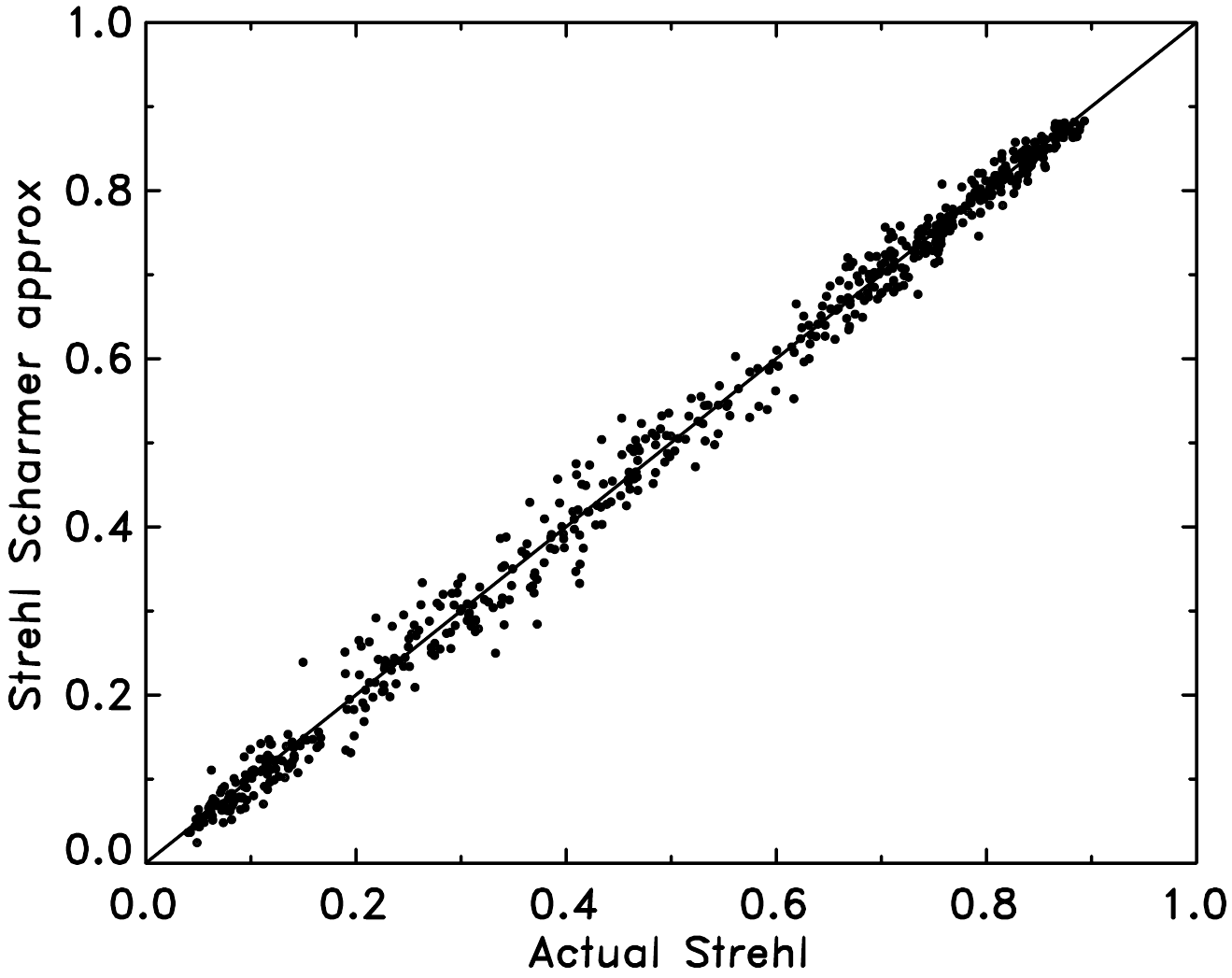 }
\includegraphics[angle=0, width=0.3\linewidth,clip] {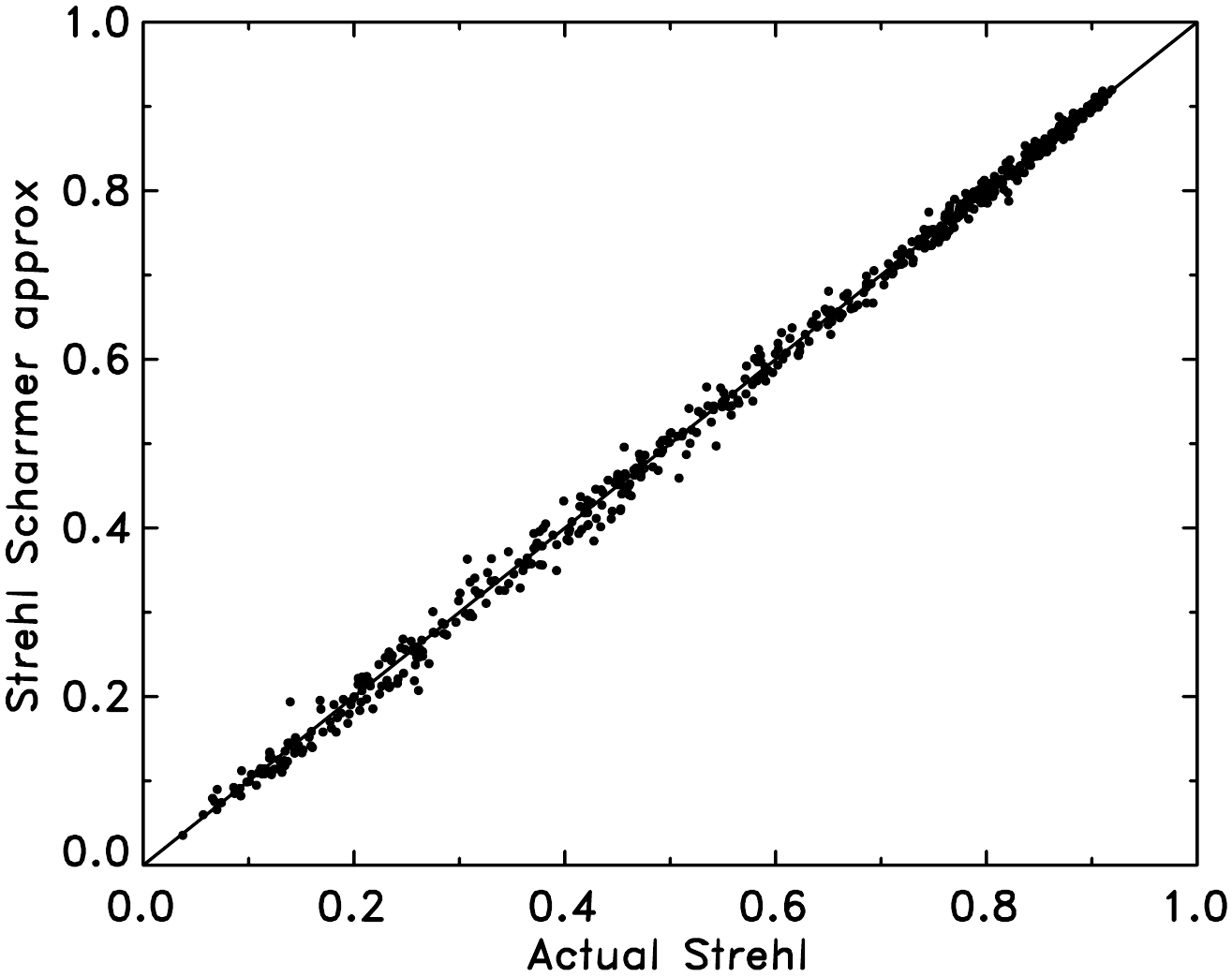 }
\includegraphics[angle=0, width=0.3\linewidth,clip] {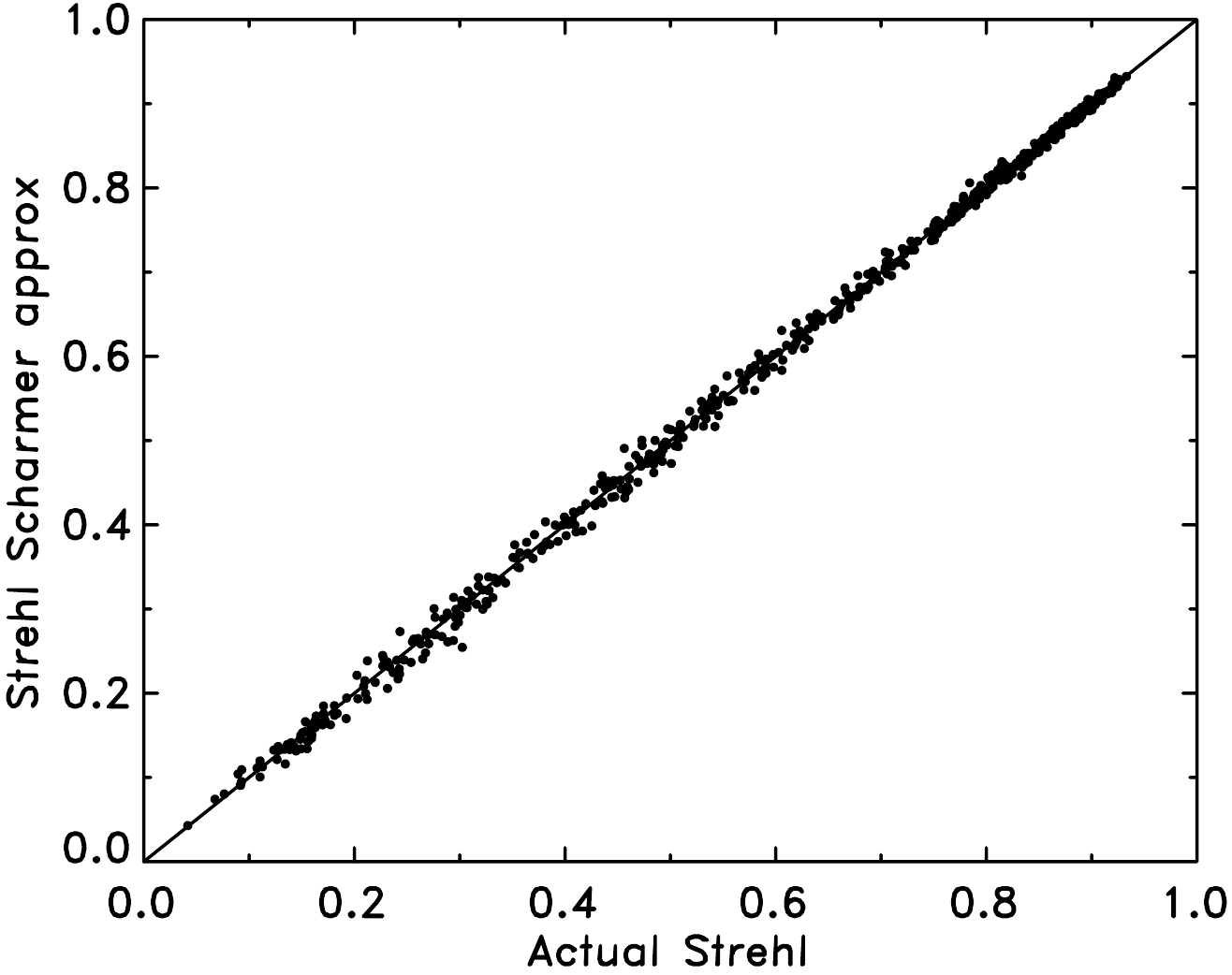 }
 \caption{Relation between actual Strehl and estimated Strehl using three different approaches: ignoring the tail of high-order aberrations (upper row), using the Löfdahl-Hillberg approach (middle row) and present approach (lower row) to compensation for the tail of high-order aberrations. Both of the latter methods assume that $r_0$ is known (from AO log files). Calculations are made with $n_1$=36 (left column), 66 (middle column) and 91 (right column) and assume that the first $n_1$ aberrations are partially compensated for with an AO system that reduces the KL amplitudes by 75\%. The range of $D/r_0$ of the input data is from 4 to 20.} 
 \label{fig:Strehl-comparison}
 \end{figure*}
 
 \subsection{Comparison of methods of straylight compensation} 
The main effect of uncompensated (by AO and/or in post-processing) high-order aberrations is to leave a PSF with a core that looks like a diffraction limited PSF with a reduced peak value, and much wider wings that may contain a large fraction of the PSF when the seeing is poor \citep[e.g., ][]{1992ESOC...42..475C}. The peak value of the PSF, normalised to that of a diffraction limited PSF, corresponds to the Strehl value, $R$, and the fraction of the PSF that is in the wings is $(1-R)$. In order to characterise the level of straylight in the wings, it is therefore sufficient to measure the peak intensity of the core of the PSF, which when normalised to its diffraction limited value equals the Strehl.  
 
Figure \ref{fig:Strehl-comparison} shows a comparison of  the Strehl values of seeing degraded and partially AO compensated PSFs, using three different values of $n_1$ (36, 66, and 91 and six different values of  $D/r_0$ in the range 4-20. Random KL coefficients from $n$=4 (not including tip-tilt terms), to $n=2016$ were used with variances given by \citet{1976JOSA...66..207N} as input, but assuming an AO system that reduced the amplitudes of KL aberrations up to $n=n_1$ by 75\%. The Strehl values were obtained by calculating the PSF and measuring its peak values, normalised to that of a diffraction limited PSF. At very low Strehl values, this corresponds to measuring the intensity of the brightest speckle.  The x-axis of the plots represents the actual Strehl of any particular sample of random KL coefficients. The y-axis gives the estimated Strehl of the same sample of low-order (up to $n=n_1$) but with different approximations of the effect of the high-order terms ($n>n_1$. The uppermost row shows the estimated Strehl values when ignoring the high-order terms (as in conventional MFBD processing without any high-order straylight compensation). The scatter of points that separate into different populations reflect the six different values of $r_0$ used to simulate the PSFs. The middle row shows the estimated Strehl values with the SD method proposed by \citet{2022A&A...668A.129L}, which is based on the use of a single random sample of high-order modes to represent the straylight. The lowermost row shows the estimated Strehl values obtained with the method proposed here - this corresponds also to the values obtained with the method of  \citet{2010A&A...521A..68S}.

The plots show that the SD method proposed by \citet{2022A&A...668A.129L} already leads to an impressive improvement in Strehl estimates, as compared to ignoring the effects of high-order aberrations (upper row). However, the lower row of plots also show that further systematic and significant improvements of the PSF is possible with the method proposed here. In Table \ref{Strehl_summary}, which builds on a total of  24,000 random samples, we quantify the improvements illustrated by the scatter plots in Fig. \ref{fig:Strehl-comparison} for three different values of $n_1$, and six different values of $D/r_0$. We also calculate the different average Strehl values of the 18 cases studied. As can be seen, the improvement of the present method \citep[and thus also the method of][since the two methods are equivalent]{2010A&A...521A..68S}, over the SD method is close to a factor $\sqrt 2$. This agrees with what could be expected intuitively, since the SD method builds on representing the high-order straylight with a single random set of KL aberrations, whereas the present method builds on calculating the average impact of a Gaussian distribution of such random KL aberrations on the straylight.

 \begin{table}[tbh]
\caption{\label{Strehl_summary} Strehl rms errors with high-order aberration compensation} 
  \centering
  \small
  \begin {tabular} {rccccccc}
    \hline
        \noalign{\smallskip}
    \mathstrut
    &&Average & Comp.& Comp. &Comp. &Ratio&\\
    $n_1$ & $D/r_0$ & Strehl & None & SD & SS & SD/SS  \\
    \hline
    \noalign{\smallskip}
    36& 3.9 & 0.84 & 0.089 & 0.016 & 0.011 & 1.42 \\ 
    & 4.9 & 0.79 & 0.121 & 0.021 & 0.015 & 1.41 \\
    & 6.5 & 0.67 & 0.177 & 0.030 & 0.021 & 1.43 \\ 
    & 9.8 & 0.46 & 0.268 & 0.039 & 0.029 & 1.37 \\ 
    & 14.0 & 0.26 & 0.327 & 0.042 & 0.030 & 1.40 \\ 
    & 19.6 & 0.10 & 0.319 & 0.031 & 0.023 & 1.37 \\ 
     \hline
     \noalign{\smallskip}
    66 & 3.9 & 0.88 & 0.051 & 0.007 & 0.005 & 1.41 \\ 
         & 4.9 & 0.83 & 0.071 & 0.010 & 0.007 & 1.41 & \\
         & 6.5 & 0.74 & 0.105 & 0.014 & 0.010 & 1.42 & \\
         & 9.8 & 0.55 & 0.166 & 0.021 & 0.015 & 1.42 & \\
        & 14.0 & 0.36 & 0.214 & 0.025 & 0.018 & 1.39 & \\
       & 19.6 & 0.18 &  0.227 & 0.025 & 0.017 & 1.41 & \\ 
  \hline
  \noalign{\smallskip}
  91 & 3.9 & 0.89 & 0.038 & 0.005 & 0.003 & 1.39 \\ 
  & 4.9 & 0.84 & 0.053 & 0.007 & 0.005 & 1.42 & \\ 
  & 6.5 & 0.76 & 0.079 & 0.009 & 0.007 & 1.43 & \\
  & 9.8 & 0.59 & 0.126 & 0.015 & 0.010 & 1.42 & \\
  & 14.0 & 0.40 & 0.167 & 0.018 & 0.013 & 1.38 & \\
  & 19.6 & 0.21 & 0.183 & 0.020 & 0.014 & 1.42 & \\

  \hline
      \noalign{\smallskip}
  \end{tabular}
  \vspace{1mm}
  \tablefoot{$n_1+1$ corresponds to the first KL aberration involved in the high-order compensation of the PSF. SD corresponds to the method of  \citet{2022A&A...668A.129L}, SS to the method proposed here, and also to that proposed by \citet{2010A&A...521A..68S}.}
  \label{Strehl_summary}
\end{table}
 
\section{Compensation for missing high-order aberrations in PD and MFBD processing}
\label{sec:comp_impl}
Within the additive white Gaussian noise model, the joint estimation of the object and the transfer functions corresponds to minimising the difference between the observed images and those modelled by the combination of the unknown object $\hat F$ and the estimated transfer functions $\hat S_k \hat T_k$,
\begin{equation}
L=\int_\nu \Bigg{\{} \sum_{k=1}^N \lvert D_k-\hat F \hat{S}_k \hat T_k \rvert ^2 \Bigg{\}}~ d\nu~,
\label{eq:21}
\end{equation}
in which $N$ is the number of images in a data-set, $\hat F=\hat F(\nu)$ is given by Eq. \ref{eq:2}, and $D_k=D_k(\nu)$, $\hat S_k=\hat S_k(\nu)$, and $\hat T_k=\hat T_k(\nu)$. Eliminating $\hat F$ results in an error metric that only contains the observed images and the unknown transfer functions. Apart from a constant, this can be written as \citep{1992JOSAA...9.1072P}
\begin{equation}
L=\int_\nu  \Bigg{\{}\sum_{j=1}^{N-1} \sum_{k=j+1}^N \lvert D_j \hat{S}_k \hat T_k - D_k \hat S_j \hat T_j \rvert ^2  \Bigg{/} \sum_{l=1}^N \lvert \hat S_l \hat T_l \rvert ^2 \Bigg{\}}~ d\nu~.  
\label{eq:22}
\end{equation}
A computationally more efficient, but less intuitive, form is \citep{1992JOSAA...9.1072P}, again apart from a constant
 \begin{equation}
L=-\int_\nu \Bigg{\{} \Bigg \lvert \sum_{k=1}^N D_k \hat S_k^* \hat T_k^* \Bigg \rvert ^2 \Bigg{/} \sum_{l=1}^N \lvert \hat S_l \hat T_l \rvert ^2  \Bigg{\}}~ d\nu~.
\label{eq:23}
\end{equation}
We first note, that if $\hat S_k$ is the same for all images in a dataset, then $\hat S_k = \hat S$ can be moved outside the summations over $j, k$ and $l$, such that $S$ in the numerator and denominator cancels out and disappears from Eqs. \ref{eq:22} and \ref{eq:23}. This means that $r_0$ cannot be determined from the observed data, but must be given as an input from an external source. In the case of SST, this source is the AO system \citep{2019A&A...626A..55S, 2024A&A...685A..32S}.

However, almost all data sets processed from SST involve wavelength scans with narrowband FPI systems \citep[CRISP2 and CHROMIS,][]{2026A&A...705A..55S}, with or without polarimetry. Such data sets are recorded during typical time intervals of up to 20~s, and usually involve substantial variations in $r_0$. Furthermore, even stable seeing conditions are associated with significant variations in the straylight contributed by the high-order aberrations and that are not captured by our approximation $\hat S_k$ - this is what explains the spread in the lowermost row of plots in Fig. \ref{fig:Strehl-comparison}. Assuming a given average $r_0$ for a given data-set, the parameter $\alpha_k$ in Eq. \ref{eq:19} can thus be used to compensate for both variations of $r_0$ within that data set, and random errors resulting from the approximations made. The extra degree of freedom provided by the parameter $\alpha_k$ can thus improve the overall fits of the data to the imaging model, thereby improving the quality of restored images and probably also reducing the number of PD or MFBD iterations. However, the set of  parameters ${\{}\alpha_k{\}}_{k=1}^N$ must be constrained by external input of $\langle r_0\rangle$ through 
\begin{equation}
\frac{1}{N} \sum_{k=1}^N \alpha_k = D/\langle r_0\rangle^{5/3}~~, \text{where}~   \langle r_0\rangle = \frac{1}{N} \sum_{k=1}^N r_{0k}~~,
\label{eq:24}
\end{equation}
or a similar constraint.

\section{Discussion} 
\label{sec:discussion}
Image reconstruction applied to short-exposure images plays a fundamental role in providing multi-wavelength diagnostics of the solar atmosphere with the highest possible image quality. PD and MFBD processing \citep{1979SPIE..207...32G, 1992JOSAA...9.1072P, 1994A&AS..107..243L, 2002SPIE.4792..146L, 2005SoPh..228..191V}, along with speckle-based techniques \citep{1970A&A.....6...85L, 1993A&A...268..374V, 2007ApOpt..46.8015W, 2008A&A...488..375W}, have proven their importance in this regard. PD and MFBD techniques are particularly well tailored to handle AO compensated images, but suffer from the limited number of aberrations that can be handled individually. 

We have earlier demonstrated that truncating the number of aberrations, without representation of the small-scale phase errors from seeing, leads to increased residual straylight. The solar granulation contrast, when compared to the true contrast obtained from simulations, therefore can be used as a measure of the Strehl of the observed or modeled PSF \citep{2019A&A...626A..55S}. This was demonstrated by comparing seeing measurements ($r_0$) obtained both with a wide-field wavefront sensor \citep{2010A&A...521A..68S} and the SST AO system \citep{2019A&A...626A..55S} to the measured granulation contrast. The top row of plots in Fig. \ref{fig:Strehl-comparison} demonstrate that ignoring high-order aberrations can result in huge errors in the Strehl values of the modeled PSFs.

We have proposed to compensate the missing high-order aberrations in PD and MFBD processing by introducing an average compensation that takes into account the collective effects of high-order seeing aberrations \citep[][see also Sect. \ref{sec:prev_work}]{2010A&A...521A..68S}, assuming that $r_0$ at the time of recording the data is known. This approach is argued for on the basis that PD and MFBD processing in practice always involves 10 or (many) more raw images, such that a compensation that is based on an average effect should work well. Simulations confirm this and show that this indeed recovers the correct contrast of solar granulation. However, the form of the compensation proposed by \citet{2010A&A...521A..68S}, with a needed averaging of many random samples of seeing, is cumbersome to combine with PD or MFBD processing, and is computationally expensive. 

\citet{2022A&A...668A.129L} proposed to simplify the implementation through a high-order compensation that only involves a single random set of high-order aberrations, with amplitudes given by statistics of turbulent seeing \citep{1976JOSA...66..207N}. Using simulations, this simpler SD approach was shown to also restore the correct contrast of solar granulation, and to offer additional advantages. First, combining image reconstruction of low-order modes with high-order compensation is simple with the SD method, and adding a single sample of given high-order aberration to the wavefront, carries no significant computational cost. Moreover, the inclusion of a single sample of random high-order aberrations seems to improve the overall convergence of the image reconstruction process. In the present work, we confirm that the SD approach indeed leads to estimated PSF's with correct Strehl values on the average (Fig. \ref{fig:Strehl-comparison}, middle row of plots), and relatively small random errors. However, the SD method has not yet been implemented for processing of real data. 

In the present work, we improve on the methods proposed by \citet{2010A&A...521A..68S} and \citet{2022A&A...668A.129L} for compensation of missing high-order aberrations. This is done by averaging the high-order part of the MTF over Gaussian distributions of KL aberrations, relying on work done by \citet{1978JOSA...68...78W}, and assuming that the accumulated contributions from many high-order aberrations are approximately uniform across the pupil. We demonstrate (Fig. \ref{fig:Strehl-comparison}, mid and lower row of plots, and Table \ref{Strehl_summary}) that this approach is significantly more accurate than that of the SD method. The mathematical expression for the proposed high-order compensation is straightforward to combine with PD and MFBD processing at very small computational cost. Allowing frame-to-frame variations of this compensation should be of particular importance, by accommodating variations of $r_0$ within any dataset. The calculations presented assume variances of high-order aberrations given by statistics of turbulent seeing, but could also include estimated variances of residual aberrations from AO compensation.

We believe that collective compensation for thousands of high-order aberrations is a more rewarding approach than individually compensating for each of these aberrations. This is suggested by the remarkably accurate Strehl estimates provided with the present method, as demonstrated in the lower row of plots in Fig. \ref{fig:Strehl-comparison}. An alternative, or complementary, approach to processing image data from 4-m solar telescopes may be to extend PD and MFBD processing to many more modes than is possible today, using wavefront sensor assisted deconvolution, as proposed by \citet{2011OExpr..19.1975J} and more recently by M. van Noort (private communication).

 \begin{acknowledgements}
The Swedish 1-m Solar Telescope is operated on the island of La Palma by the Institute for Solar Physics of Stockholm University in the Spanish Observatorio del Roque de los Muchachos of the Instituto de Astrofísica de Canarias. The Institute for Solar Physics is supported by a grant for research infrastructures of national importance from the Swedish Research Council (registration number 2021-00169). The European Solar Telescope project is supported by a grant for research infrastructures from the Swedish Research Council (registration number 2023-00169). Mats Löfdahl is thanked for providing access to calculated Karhunen--Lo\`eve polynomials, and comments on the manuscript.
\end{acknowledgements}


 \end{document}